\documentclass[sigconf]{acmart}
\usepackage{enumitem}
\usepackage{multirow}
\usepackage{makecell}
\usepackage{graphicx}
\usepackage{xcolor}
\usepackage{colortbl}

\copyrightyear{2025}
\acmYear{2025}
\setcopyright{cc}
\setcctype{by}
\acmConference[CIKM '25]{Proceedings of the 34th ACM International Conference on Information and Knowledge Management}{November 10--14, 2025}{Seoul, Republic of Korea}
\acmBooktitle{Proceedings of the 34th ACM International Conference on Information and Knowledge Management (CIKM '25), November 10--14, 2025, Seoul, Republic of Korea}
\acmISBN{979-8-4007-2040-6/2025/11}
\acmDOI{10.1145/3746252.3761517}

\begin{document}

\title{THEME: Enhancing Thematic Investing with Semantic Stock Representations and Temporal Dynamics}

\author{Hoyoung Lee}
\authornote{Equal contribution}
\affiliation{
\institution{Ulsan National Institute of Science and Technology}
\state{Ulsan}
\country{Republic of Korea}
}
\email{hoyounglee@unist.ac.kr}

\author{Wonbin Ahn}
\authornotemark[1] % Equal contribution
\affiliation{
\institution{LG AI Research}
\state{Seoul}
\country{Republic of Korea}
}
\email{wonbin.ahn@lgresearch.ai}

\author{Suhwan Park}
\affiliation{
\institution{Ulsan National Institute of Science and Technology}
\state{Ulsan}
\country{Republic of Korea}
}
\email{suhwan@unist.ac.kr}

\author{Jaehoon Lee}
\affiliation{
\institution{LG AI Research}
\state{Seoul}
\country{Republic of Korea}
}
\email{jaehoon.lee@lgresearch.ai}

\author{Minjae Kim}
\affiliation{
\institution{LG AI Research}
\state{Seoul}
\country{Republic of Korea}
}
\email{minjae.kim@lgresearch.ai}

\author{Sungdong Yoo}
\affiliation{
\institution{LG AI Research}
\state{Seoul}
\country{Republic of Korea}
}
\email{sungdong.yoo@lgresearch.ai}

\author{Taeyoon Lim}
\affiliation{
\institution{LG AI Research}
\state{Seoul}
\country{Republic of Korea}
}
\email{taeyoon.lim@lgresearch.ai}

\author{Woohyung Lim}
\authornote{Corresponding author}
\affiliation{
\institution{LG AI Research}
\state{Seoul}
\country{Republic of Korea}
}
\email{w.lim@lgresearch.ai}

\author{Yongjae Lee}
\authornotemark[2] % Corresponding author
\affiliation{
\institution{Ulsan National Institute of Science and Technology}
\state{Ulsan}
\country{Republic of Korea}
}
\email{yongjaelee@unist.ac.kr}

\renewcommand{\shortauthors}{Lee and Ahn et al.}

\begin{abstract}
Thematic investing, which aims to construct portfolios aligned with structural trends, remains a challenging endeavor due to overlapping sector boundaries and evolving market dynamics. A promising direction is to build semantic representations of investment themes from textual data. However, despite their power, general-purpose LLM embedding models are not well-suited to capture the nuanced characteristics of financial assets, since the semantic representation of investment assets may differ fundamentally from that of general financial text. To address this, we introduce THEME, a framework that fine-tunes embeddings using hierarchical contrastive learning. THEME aligns themes and their constituent stocks using their hierarchical relationship, and subsequently refines these embeddings by incorporating stock returns. This process yields representations effective for retrieving thematically aligned assets with strong return potential. Empirical results demonstrate that THEME excels in two key areas. For thematic asset retrieval, it significantly outperforms leading large language models. Furthermore, its constructed portfolios demonstrate compelling performance. By jointly modeling thematic relationships from text and market dynamics from returns, THEME generates stock embeddings specifically tailored for a wide range of practical investment applications.
\end{abstract}





\maketitle

\section{Introduction}

\begin{figure}[htbp]
  \centering
  \includegraphics[width=1.0\linewidth]{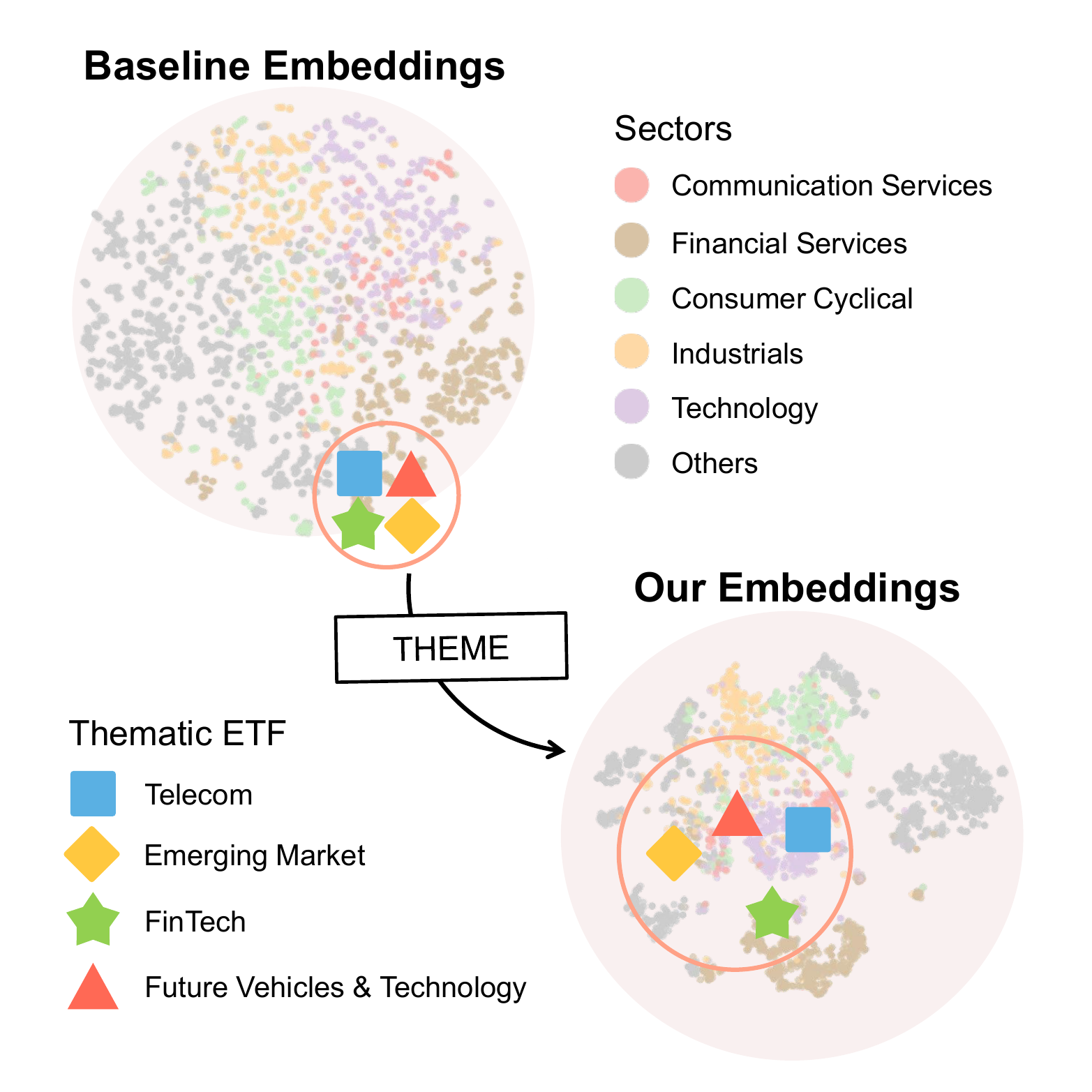}
\caption{A t-SNE comparison of stock embeddings before and after applying THEME. (Upper Left) The baseline model fails to group stocks by investment theme. (Lower Right) In contrast, our tuned embeddings form distinct, meaningful clusters corresponding to investment themes like FinTech and Future Vehicles, significantly improving interpretability.}
  \label{fig:tsne-comparison}
\end{figure}

\begin{figure*}[htbp]
  \centering
  \includegraphics[width=\linewidth]{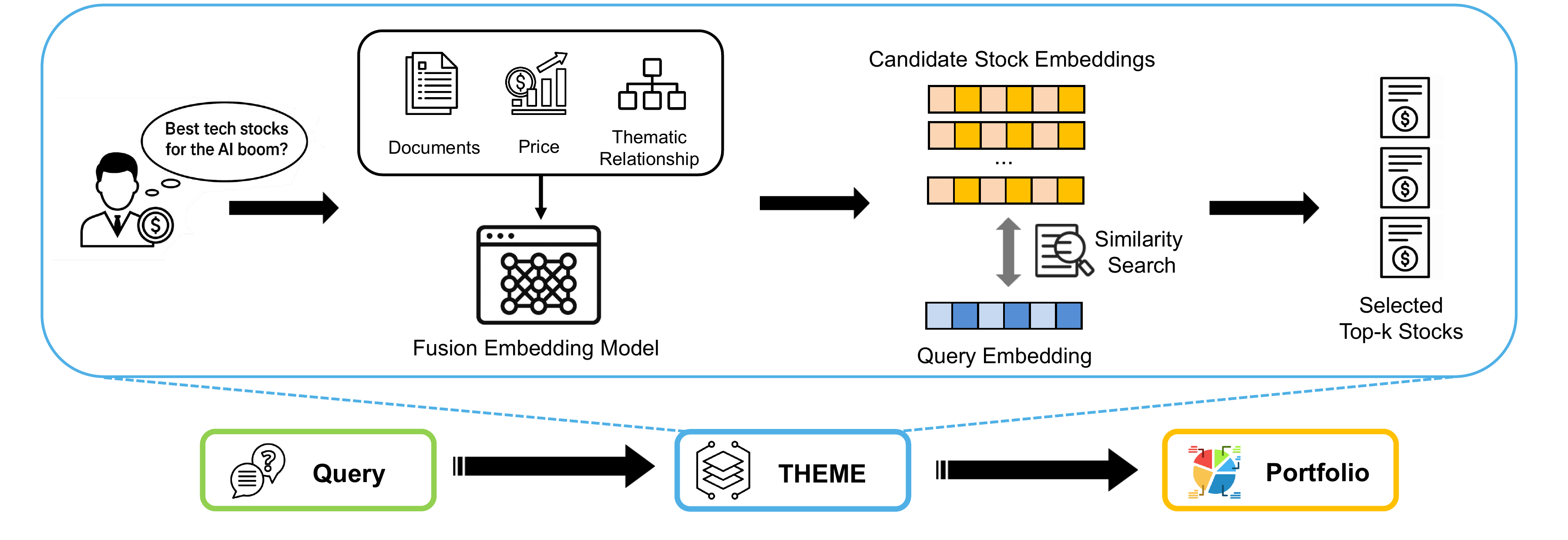}
  \vspace{-2em}
  \caption{
  Overview of how the THEME system is used in practice and how it works. Given a user-provided thematic description—such as "AI software and chipmakers"—THEME semantically embeds the input and retrieves stocks that are both thematically aligned and investment-suitable. These selected stocks can then be used to construct a portfolio for real-world investment.
  }
  \label{fig:proc}
  \vspace{-1em}
\end{figure*}

Thematic investing is a popular strategy that constructs portfolios around structural trends such as artificial intelligence, renewable energy, or cybersecurity. Unlike sector-based investing, thematic strategies are inherently cross-sectoral and dynamic: relevant companies often span diverse industries and change over time in response to innovation, regulation, and market sentiment.

Despite its appeal, current thematic investment methods rely heavily on static ETF compositions or expert-defined lists. These approaches lack adaptability and are often slow to reflect emerging companies or shifts in thematic relevance. As a result, investors may miss out on timely opportunities or hold outdated portfolios that no longer represent the target theme.

To address this gap, we propose THEME, a scalable system for thematic stock retrieval and portfolio construction. As illustrated in Figure~\ref{fig:tsne-comparison}, traditional embedding models trained on general-purpose text fail to produce meaningful clusters of thematically related stocks. This highlights a key limitation: financial texts contain domain-specific semantics that general embeddings cannot fully capture. Our method addresses this by producing domain-tuned embeddings that reflect thematic structure, enabling more interpretable and effective retrieval.

Figure~\ref{fig:proc} shows how THEME operates in practice. Given a user-defined thematic query, THEME encodes the theme, retrieves semantically and financially relevant stocks, and returns candidates suitable for portfolio construction. The system integrates seamlessly into real-world workflows, supporting both discretionary screening and systematic portfolio updates.

At the core of THEME is a hierarchical contrastive learning framework that combines two sources of supervision. First, semantic signals derived from financial text (e.g., business descriptions, regulatory filings, and news) provide structural thematic alignment. Second, temporal signals based on recent returns guide the model to favor stocks with short-term investment relevance. This two-stage process enables the model to capture both long-term thematic meaning and dynamic market responsiveness.

To support robust learning and generalization, we also construct the \textit{Thematic Representation Set (TRS)}—a dataset that extends beyond real-world ETF holdings by incorporating sectoral taxonomies and news-derived themes. TRS enables THEME to support hundreds of broad and niche themes, including those not covered by existing ETF products.

In summary, we make the following contributions:

\begin{itemize}[topsep=2.5pt, itemsep=2.5pt, leftmargin=1.0em]
    \item \textbf{A unified semantic-temporal framework for theme modeling:}  
    We propose a contrastive learning approach that embeds stocks using both textual descriptions and recent return behavior. This enables theme-aware retrieval that captures both structural meaning and short-term responsiveness.

    \item \textbf{A domain-tuned embedding space for finance:}  
    We show that general-purpose embeddings fail to organize stocks by thematic relevance. In contrast, our fine-tuned embeddings yield clearer semantic structure, improving retrieval and interpretability.

    \item \textbf{A hierarchical modeling strategy for thematic structure:}  
    Our method supports a wide range of themes, from broad categories to niche subthemes, by applying a multi-level contrastive framework that reflects the layered nature of thematic semantics.

    \item \textbf{A scalable thematic dataset (TRS):}  
    We construct the TRS, a large-scale stock-theme dataset that extends beyond ETF coverage by incorporating industry taxonomies and financial news. This improves coverage of emerging and underrepresented themes.

    \item \textbf{Real-world evaluation for portfolio construction:}  
    We implement and assess a full-system pipeline that improves retrieval precision and portfolio performance across multiple model backbones. The system is compatible with both discretionary and systematic investment workflows.
\end{itemize}

\section{Related Works}
The rise of thematic investing in recent years has prompted research into systematic methods for identifying theme-relevant companies. A key challenge is that themes often cut across conventional industry boundaries. Prior approaches have typically relied on expert curation or static industry classifications, which struggle to capture the full breadth of relevant firms. To address this, researchers have increasingly turned to natural language processing (NLP) techniques applied to financial text.

For instance, Rao et al.\ \cite{rao2023thematicnlp} demonstrate that \emph{fit-for-purpose} language models can effectively associate companies with long-term themes by comparing thematic keywords to business descriptions. Similarly, Vamvourellis et al.\ \cite{vamvourellis2023secembeddings} show that embeddings derived from SEC filings using fine-tuned transformer models not only reproduce standard industry groupings but also reveal new cross-sector relationships. These semantic embeddings have enabled applications such as clustering companies by thematic exposure and constructing thematic baskets. Building on this line of work, Takayanagi et al.\ \cite{takayanagi2024setn} introduce a hybrid approach (SETN) that incorporates both textual and network-based signals to enhance stock representations for thematic investing.

However, financial markets are inherently non-stationary, and a company’s thematic alignment may shift over time due to structural changes, strategic pivots, or macroeconomic developments \cite{li2023temporal}. Traditional static models fail to capture this temporal dynamic. In response, recent work has explored time-aware representations. For example, Hwang et al.\ \cite{hwang_simstock_2023} propose SimStock, a self-supervised framework that generates temporally-sensitive stock embeddings, improving performance on tasks such as thematic index tracking. In practice, dynamic thematic strategies are also gaining popularity, frequently updating exposures based on short-term signals such as news sentiment and price momentum \cite{gu2020empirical}.

Meanwhile, large language models (LLMs) like GPT-4 have shown growing utility in financial applications \cite{lee2023overview, kim2023llms, nie2024survey, lee2025your, lee2024overview}. In theory, these models can infer thematic relevance from contextual knowledge alone \cite{openai2023gpt4}. Yet, general-purpose LLMs often fall short in this domain: they may lack access to up-to-date disclosures or fail to detect subtle, domain-specific quantitative signals, leading to inconsistent or incomplete outputs \cite{yang2023finllm}. BloombergGPT \cite{wu2023bloomberggpt} highlights the potential of domain-specific models, but its application to thematic stock selection remains unexplored.

Despite these advances, no existing system fully integrates semantic NLP techniques with explicit temporal modeling to support dynamic thematic investing \cite{wang2023kgtransformer}. This motivates the development of THEME, a unified framework that combines language-based stock embeddings with time-aware performance modeling. By capturing both the textual signals of thematic relevance and their evolution over time, THEME aims to provide a practical and robust solution for identifying and tracking theme-aligned stocks in a non-stationary market environment.

\begin{figure*}[htbp]
  \centering
  \includegraphics[width=\linewidth]{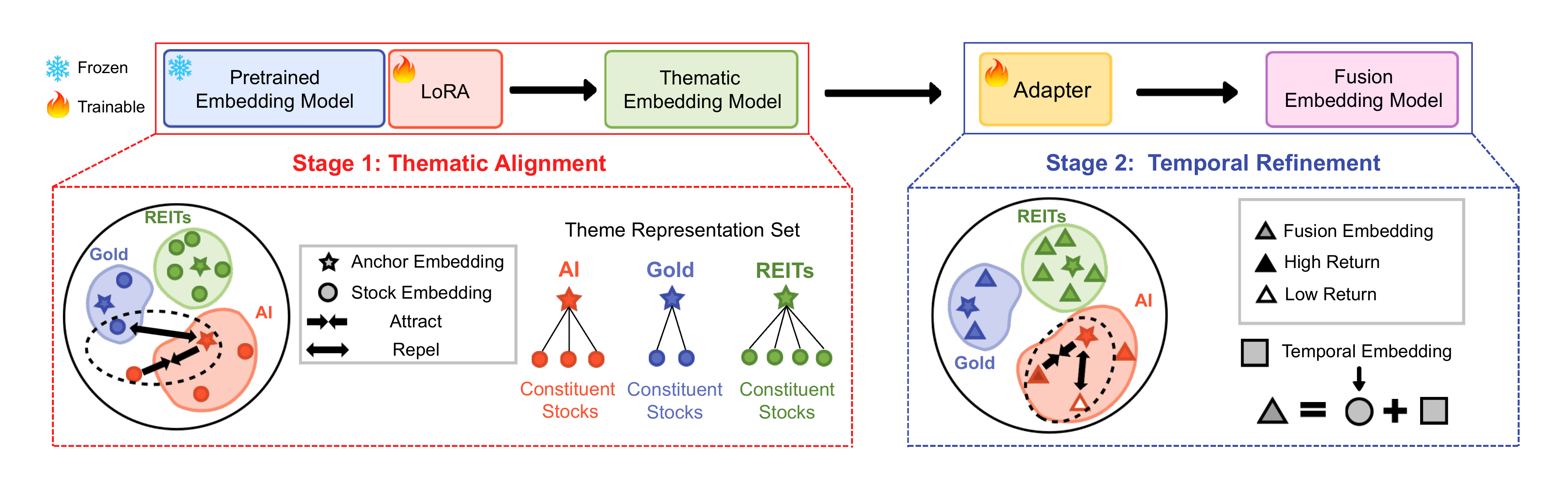}
  \vspace{-2em}
\caption{
Overview of the two-stage hierarchical contrastive learning framework. 
By leveraging the explicit relationship between a theme and its known constituent stocks as the primary training signal, the framework first learns semantic embeddings via contrastive learning. 
Subsequently, a lightweight adapter refines these semantic embeddings by fusing them with temporal embeddings derived from recent stock returns. 
The resulting model generates theme-aware stock representations that jointly capture semantic meaning and temporal signals, suitable for dynamic portfolio construction.
}
  \label{fig:method}
\end{figure*}

\section{Methodology}

This work introduces a hierarchical contrastive learning framework for thematic investing (Figure \ref{fig:method}). This framework generates stock embeddings that capture both semantic alignment with theme descriptions and sensitivity to short-term return patterns. Prior contrastive methods often focus solely on static textual similarity. In contrast, our approach jointly models semantic meaning and temporal signals to support dynamic, theme-aware stock retrieval. It is designed to scale to thousands of stocks across asset classes and is optimized for real-world deployment.

We refer to this framework as \textbf{hierarchical contrastive learning} for two key reasons. First, it directly utilizes the hierarchical relationship between themes and their individual constituent stocks in the learning process. Second, it applies two levels of contrastive objectives hierarchically. The first stage is a semantic objective that aligns stocks with the conceptual meaning of a theme, followed sequentially by a temporal objective that refines the embeddings using recent return behavior. This multi-layered hierarchical structure enables the progressive refinement of stock representations from static thematic semantics to dynamic investment relevance.

To support this two-stage learning process, we construct the TRS. TRS begins with 1,153 real-world thematic ETFs, each linked to a textual description and a list of constituent stocks. While ETFs offer strong supervision signals, their coverage is incomplete and biased toward trending sectors such as IT and clean energy. To address this, we incorporate sector and industry classification systems to expand our theme universe to approximately 200 unique themes.

Each TRS record contains a theme label, a textual summary, and its associated list of stocks. For each stock, we construct a rich textual profile by aggregating data from SEC filings and financial news, allowing stocks to participate in multiple themes. This contrasts with traditional systems where each stock belongs to a single sector. Moreover, TRS is updated dynamically with live information, ensuring continued relevance as new themes emerge.

\subsection{Stage 1: Thematic Alignment}
In the first stage, we learn a shared embedding space that semantically aligns the textual profiles of stocks with the descriptions of themes. This process aims to position semantically related stocks and themes closer together within the embedding space.

First, we use a large-scale, pre-trained text embedding model with frozen parameters as the backbone, denoted by $f_{\Phi}(\cdot)$, where $\Phi$ represents the pre-trained weights. We apply LoRA (Low-Rank Adaptation) \cite{hu2022lora} to this backbone model, introducing a small set of trainable parameters, $\theta$. The final, combined model is denoted as $f'_{\theta}(\cdot)$,  representing the LoRA-adapted backbone $f_{\Phi}(\cdot)$.

Given a set of theme descriptions $\mathcal{T} = \{t_1, t_2, \dots, t_M\}$ and a set of stock textual profiles $\mathcal{S} = \{s_1, s_2, \dots, s_N\}$, the final model $f'_{\theta}(\cdot)$ is used to transform each text into a $d$-dimensional vector:
\begin{equation}
\label{eq:embeddings}
z_i = f'_{\theta}(t_i) \in \mathbb{R}^d, \quad h_j = f'_{\theta}(s_j) \in \mathbb{R}^d
\end{equation}
where $z_i$ is the embedding for a theme $t_i \in \mathcal{T}$, and $h_j$ is the embedding for a stock $s_j \in \mathcal{S}$.

The learning process uses each theme embedding $z_i$ as an anchor to pull closer the embedding of a constituent stock $h_j^+$ from the theme, which serves as the positive pair, while pushing away the embeddings of non-constituent stocks $h_j^-$ as negative pairs.

The semantic alignment loss $\mathcal{L}_{\text{align}}$ is given by:
\begin{equation}
\label{eq:loss_function}
\mathcal{L}_{\text{align}} = -\log \frac{\exp(\text{sim}(z_i, h_j^+)/\tau)}{\exp(\text{sim}(z_i, h_j^+)/\tau) + \sum_{k} \exp(\text{sim}(z_i, h_j^-)/\tau)}
\end{equation}

Here, sim$(\cdot, \cdot)$ is cosine similarity and $\tau$ is the temperature hyperparameter. The efficiency of LoRA enables the rapid and dynamic update of the resulting semantic embeddings to reflect new themes or evolving semantic contexts.

\subsection{Stage 2: Temporal Refinement}
The second stage refines the semantic embeddings by incorporating dynamic, short-term return signals. To achieve this, we introduce a lightweight 2-layer adapter, denoted as $\mathcal{A}_{\phi}$, with trainable parameters $\phi$. This adapter is designed to fuse semantic context with dynamic temporal patterns. It takes two distinct inputs for a stock $s_j$: its semantic embedding ${h}_j$ from Stage 1, and its past $L$ trading days of daily returns, ${r}_j$, where $L$ is the lookback period. The adapter jointly processes these inputs to produce a single, fusion embedding ${h}'_j$:
\begin{equation}
\label{eq:temporally_refined_embedding}
{h}'_j = \mathcal{A}_{\phi}({h}_j, {r}_j)
\end{equation}

The adapter's parameters $\phi$ are optimized via a triplet loss function designed to rank stocks within a theme based on future returns. For a given theme $t_i$, we first select a positive sample $s_p$ and a negative sample $s_n$ from its constituent stocks. The stock with the higher forward return over a horizon of $H$ days is designated as the positive sample, while the other serves as the negative. These samples, along with the theme's semantic embedding $z_i$ as an anchor, form a training triplet $(z_i, h'_p, h'_n)$, where $h'_p$ and $h'_n$ are the fusion embeddings of the positive and negative stocks, respectively. The training objective is to minimize the following loss, thereby ensuring that the anchor is closer in the embedding space to the positive sample than to the negative sample by at least a margin $m$:
\begin{equation}
\label{eq:triplet_loss_final_fused}
\mathcal{L}_{\text{triplet}} = \left[ \text{sim}\left({z}_i, {h}'_n\right) - \text{sim}\left({z}_i, {h}'_p\right) + m \right]_+
\end{equation}
This approach allows the model to learn a nuanced representation that is not only thematically relevant but also sensitive to recent market dynamics, all while preserving the rich knowledge in the pre-trained semantic embeddings. Additional training details are provided in Section~\ref{sec:experiments}.

\subsection{Stage 3: Inference Pipeline}
At inference time, a user query $q$ is encoded into a query embedding $z_q$ via a semantic model $f'_{\theta}(\cdot)$. We perform retrieval by ranking a corpus of pre-computed stock embeddings ${h}'_{j}$ based on their cosine similarity to the query vector $z_q$. The resulting top-$K$ list is subsequently utilized for downstream financial applications, including thematic screening, portfolio construction, and index design.

\subsection{System Implementation and Integration}

Our system is implemented as a modular, cloud-native application. Embedding and similarity computation scale linearly with the number of stocks, enabling fast inference across global universes. REST APIs expose the core functionality, supporting integration with discretionary research platforms, automated portfolio engines, and personalized investing tools. Feedback mechanisms allow iterative refinement based on user inputs.

The platform is also extensible to real-time signals such as ESG events, earnings announcements, or patent activity. These extensions enable advanced use cases including event-driven rebalancing, theme-based risk monitoring, and semantic pre-filtering for financial NLP pipelines.

\begin{table*}[t]
\centering
\small

\caption{
\textbf{Comparison of retrieval performance across various methods before and after applying THEME.} 
Metrics include Hit Rate (HR) and Precision (P) at $k \in \{3, 5, 10\}$, where HR@$k$ denotes the fraction of queries for which at least one relevant item appears in the top-$k$ results, and P@$k$ measures the proportion of relevant items among the top-$k$ retrieved.
}
\label{tab:exp1_retrieval}
\renewcommand{\arraystretch}{0.9}
\setlength{\tabcolsep}{3pt} 
\vspace{-1em}
\resizebox{\textwidth}{!}{
\begin{tabular}{ll|cc|cc|cc|cc|cc|cc}
\toprule
\multirow{2}{*}{\textbf{Model}} & \multirow{2}{*}{\textbf{Size}} & \multicolumn{2}{c|}{\textbf{HR@3}} & \multicolumn{2}{c|}{\textbf{P@3}} & \multicolumn{2}{c|}{\textbf{HR@5}} & \multicolumn{2}{c|}{\textbf{P@5}} & \multicolumn{2}{c|}{\textbf{HR@10}} & \multicolumn{2}{c}{\textbf{P@10}} \\
& & Vanilla & Ours & Vanilla & Ours & Vanilla & Ours & Vanilla & Ours & Vanilla & Ours & Vanilla & Ours \\
\midrule
\rowcolor{gray!10}
voyage-2-finance~\cite{voyageai2025} & Unknown & 0.4278 & - & 0.2285 & - & 0.4742 & - & 0.2134 & - & 0.5567 & - & 0.1948 & - \\
\rowcolor{gray!10}
Fin-E5~\cite{fine5} & Unknown & 0.4948 & - & 0.3247 & - & 0.5515 & - & 0.3113 & - & 0.6289 & - & 0.2835 & - \\
\rowcolor{yellow!10}
GPT-4.1~\cite{openai2023gpt4} & Unknown & 0.7113 & - & 0.5189 & - & 0.7731 & - & 0.4505 & - & 0.8092 & - & 0.3690 & - \\
\rowcolor{yellow!10}
Gemini-2.5~\cite{gemini} & Unknown & 0.6494 & - & 0.4020 & - & 0.6649 & - & 0.3463 & - & 0.7268 & - & 0.3190 & - \\
bge-small-en-v1.5~\cite{bge_embedding} & 33M & 0.1392 & 0.6031 & 0.0584 & 0.3557 & 0.1701 & 0.7423 & 0.0495 & 0.3722 & 0.2577 & 0.8247 & 0.0541 & 0.3701 \\
bilingual-embedding-large~\cite{conneau2019unsupervised} & 559M & 0.1598 & 0.6443 & 0.0687 & 0.3935 & 0.1959 & 0.7371 & 0.0691 & 0.3918 & 0.3041 & 0.8093 & 0.0686 & 0.3825 \\
multilingual-e5-large-instruct~\cite{wang2024multilingual} & 560M & 0.1753 & 0.6959 & 0.0704 & 0.4588 & 0.2577 & 0.7732 & 0.0773 & 0.4392 & 0.3505 & 0.8660 & 0.0789 & 0.4206 \\
stella\_en\_1.5B\_v5~\cite{zhang2025jasperstelladistillationsota} & 1.5B & 0.2474 & 0.7320 & 0.1375 & 0.5275 & 0.3557 & 0.8299 & 0.1412 & 0.5237 & 0.4536 & 0.8814 & 0.1387 & 0.4856 \\
gte-Qwen2-1.5B-instruct~\cite{li2023towards} & 1.5B & 0.3144 & 0.6186 & 0.1684 & 0.4124 & 0.3763 & 0.6959 & 0.1608 & 0.4031 & 0.4639 & 0.8093 & 0.1464 & 0.3907 \\
SFR-Embedding-Mistral~\cite{SFRAIResearch2024} & 7B & 0.3711 & 0.7887 & 0.2062 & 0.6082 & 0.4639 & 0.8557 & 0.2052 & 0.5845 & 0.5567 & 0.9124 & 0.1995 & 0.5619 \\
GritLM-7B~\cite{muennighoff2024generative} & 7B & 0.0825 & \textbf{0.8196} & 0.0344 & 0.6014 & 0.1237 & 0.8814 & 0.0371 & 0.5907 & 0.1753 & 0.9124 & 0.0320 & \textbf{0.5701} \\
gte-Qwen2-7B-instruct~\cite{li2023towards} & 7B & 0.5206 & 0.7938 & 0.3299 & 0.5790 & 0.6031 & 0.8351 & 0.3227 & 0.5680 & 0.7165 & 0.9072 & 0.2985 & 0.5366 \\
e5-mistral-7b-instruct~\cite{wang2023improving} & 7B & 0.3454 & 0.7887 & 0.1770 & 0.5962 & 0.4691 & 0.8763 & 0.1907 & 0.5794 & 0.5412 & \textbf{0.9330} & 0.1912 & 0.5552 \\
Linq-Embed-Mistral~\cite{linq} & 7B & 0.5155 & \textbf{0.8196} & 0.3522 & \textbf{0.6289} & 0.5773 & \textbf{0.8918} & 0.3340 & \textbf{0.6041} & 0.6546 & 0.9278 & 0.3155 & \textbf{0.5701} \\
\bottomrule
\end{tabular}
}
\end{table*}

\section{Experiments}\label{sec:experiments}
We design a series of experiments to evaluate the effectiveness of our proposed system, THEME, in both retrieval quality and investment utility. First, we evaluate how accurately the model retrieves stocks relevant to a given theme. This measures the semantic alignment between theme descriptions and candidate stocks, using metrics such as Hit Rate (HR) and Precision (P) at various cutoff thresholds. Second, we assess whether the retrieved stocks lead to improved investment outcomes. We construct portfolios based on the top-ranked results and compare their performance against those generated by baseline methods. Third, we perform an ablation study to analyze the role of theme representation. In particular, we compare two anchor strategies: (i) using a group of representative stocks, and (ii) using a textual theme description.

\subsection{Experimental Setup}
We train the model using a two-tier dataset. The first component consists of 1,153 real-world thematic ETFs covering approximately 3,000 unique U.S. equities. To ensure quality, we filter for ETFs with at least 10 constituents. The remaining 969 ETFs are randomly split into training (678), validation (97), and test (194) sets. While the validation and test sets are preserved for real-world evaluation, the training data is augmented via our TRS methodology to a final set of 196 themes. This allows the model to learn from a diverse thematic universe while being tested on true market data.
For the temporal refinement stage, we use two years of historical U.S. market data, generating training samples via a rolling window with a lookback period of $L=60$ trading days. Pairs of stocks within each theme are labeled based on their relative forward returns over a horizon of $H=14$ days. 

\subsection{Experiment 1: Retrieval Performance}
\label{sec:experiment1}
In this experiment, we evaluate the retrieval performance of THEME against various base models. The evaluation includes several types of models: domain-specific embedding models (gray rows), state-of-the-art LLMs (yellow rows), and general-purpose embedding models. Each model ranks a universe of stocks based on a given theme description, and we measure the retrieval quality using Hit Rate (HR) and Precision (P) at $k \in \{3, 5, 10\}$.

The results in Table~\ref{tab:exp1_retrieval} clearly show that THEME dramatically improves retrieval performance. For example, applying THEME to our \texttt{Linq-Embed-Mistral} model boosted its HR@3 from 0.5155 to 0.8196 and P@3 from 0.3522 to 0.6289. Similar gains were observed across other 7B-scale models. Notably, this enhancement was not limited to large models. Even the small \texttt{bge-small-en-v1.5} model, when combined with THEME, surpassed larger vanilla 7B models and the domain-specific embedding models.

Furthermore, models enhanced with THEME recorded superior performance compared to powerful SOTA LLMs such as \texttt{GPT-4.1} and \texttt{Gemini-2.5}. This highlights that while general-purpose LLMs possess a vast amount of knowledge, a more specialized strategy of integrating theme-aligned supervision is ultimately more effective for this specific task. Ultimately, the top performance of models like \texttt{Linq-Embed-Mistral} and \texttt{GritLM-7B} confirms that THEME is an effective and model-agnostic methodology for significantly enhancing thematic stock retrieval, regardless of the underlying model's size or type.

\begin{table*}[t]
\small
\centering
\caption{
\textbf{Model performance comparison across key financial metrics.} 
Each column reports performance at different top-$k$ cutoffs ($k \in \{3, 5, 10\}$), with Sharpe Ratio (SR) and Cumulative Return (CR) indicating profitability (higher is better, $\uparrow$), and Maximum Drawdown (MDD) indicating risk (lower is better, $\downarrow$).
}
\label{tab:performance}
\vspace{-1em}
\setlength{\tabcolsep}{8pt} 
\renewcommand{\arraystretch}{0.9}
\resizebox{\textwidth}{!}{
\begin{tabular}{l l | c c c | c c c | c c c}
\toprule
\textbf{Model} & \textbf{Type} 
& \textbf{SR@3 } & \textbf{SR@5 } & \textbf{SR@10 } 
& \textbf{MDD@3 } & \textbf{MDD@5 } & \textbf{MDD@10 } 
& \textbf{CR@3 } & \textbf{CR@5 } & \textbf{CR@10 } \\
\midrule
\multirow{2}{*}{Linq-Embed-Mistral}  
& Vanilla & 0.4870 & 0.4530 & 0.4991 & -0.2551 & -0.2618 & -0.2534 & 0.0907 & 0.0831 & 0.0938 \\
& Ours    & 0.5881 & 0.5913 & 0.5432 & -0.2526 & -0.2474 & -0.2440 & 0.1187 & 0.1176 & 0.1043 \\
\midrule
\multirow{2}{*}{gte-Qwen2-7B-instruct}  
& Vanilla & 0.5014 & 0.4293 & 0.4712 & -0.2427 & -0.2413 & -0.2392 & 0.0917 & 0.0749 & 0.0843 \\
& Ours    & \textbf{0.7592} & \textbf{0.7711} & \textbf{0.6893} & \textbf{-0.2378} & \textbf{-0.2320} & \textbf{-0.2338} & \textbf{0.1645} & \textbf{0.1650} & \textbf{0.1422} \\
\midrule
\multirow{2}{*}{GritLM-7B}  
& Vanilla & 0.5744 & 0.5912 & 0.6324 & -0.2431 & -0.2409 & -0.2407 & 0.1154 & 0.1169 & 0.1273 \\
& Ours    & 0.5952 & 0.5163 & 0.5291 & -0.2683 & -0.2589 & -0.2463 & 0.1196 & 0.0981 & 0.1007 \\
\bottomrule
\end{tabular}}
\flushleft{\footnotesize \textit{Note: Average performance of real-world Thematic ETFs: SR 0.4845, MDD -0.2368, CR 0.0672.}}
\end{table*}

\subsection{Experiment 2: Portfolio Construction}
\label{sec:experiment2}
Having confirmed that THEME retrieves relevant stocks (Experiment \ref{sec:experiment1}), we now assess their investment performance. To this end, we evaluate the performance of portfolios constructed using the top-$K$ stocks selected by  THEME over a rolling test period from April 23, 2024 to April 29, 2025.

For each window, we construct equal-weighted portfolios using the top-$K$ stocks ranked by similarity to the theme anchor. The average daily return of these $K$ stocks is then recorded over the subsequent 14 trading days. Chaining the returns across all windows forms a continuous daily return series over the entire test period.

Portfolio performance is evaluated with the following metrics:
\begin{itemize}
    \item \textbf{Cumulative Return (CR):}
    \[
    \mathrm{CR}_t = \prod_{i=1}^{t} (1 + r_i) - 1,
    \]
    where \( r_i \) is the portfolio return on day \( i \).
    
    \item \textbf{Sharpe Ratio (SR):}
    \[
    \mathrm{SR} = \frac{\bar{r}}{\sigma_r} \cdot \sqrt{252},
    \]
    where \( \bar{r} \) and \( \sigma_r \) are the mean and standard deviation of daily returns.

    \item \textbf{Maximum Drawdown (MDD):}
    \[
    \mathrm{MDD} = \min_t \left( \frac{V_t}{\max_{s \leq t} V_s} - 1 \right),
    \]
    with \( V_t \) representing cumulative portfolio value at time \( t \).
\end{itemize}
As summarized in Table~\ref{tab:performance}, THEME consistently improves portfolio quality over vanilla approaches. The performance gains are clear: for the \texttt{gte-Qwen2-7B-instruct}, SR@3 increases from 0.5014 to 0.7592, CR@3 from 0.0917 to 0.1645, and MDD improves from –0.2427 to –0.2378. These results also outperform the baseline performance of investing in the real thematic ETFs, which achieves a Sharpe ratio of 0.4845, CR of 0.0672, and MDD of –0.2368. Similar advantages are observed across other $K$ values and model variants.

Notably, the impact of THEME varies depending on the baseline's initial strength. On a powerful baseline like \texttt{GritLM-7B}, which performs exceptionally well in its vanilla state, our method offers marginal gains in specific top-3 scenarios while not consistently surpassing it across all metrics. This suggests that our temporal refinement provides the most significant lift to models that are not yet fully optimized for the task, such as \texttt{gte-Qwen2-7B-instruct}. Ultimately, these findings confirm that fusing semantic and temporal signals is a highly effective strategy for creating profitable and resilient portfolios.

\subsection{Experiment 3: Ablation on Anchor and Dataset Strategy}
\label{sec:experiment3}
We conduct an ablation study to investigate how two key aspects of our contrastive learning setup affect retrieval quality: (i) the choice of anchor during training, and (ii) the composition of the training dataset. Each factor is evaluated independently using a consistent set of backbone models and evaluation metrics.

\begin{table}[htbp]
\small
\centering
% \caption{Ablation study comparing different anchor selection strategies for semantic alignment. In our setting, the anchor is the description of a thematic ETF and the positive is one of its constituent stocks. In the counterpart setting, both the anchor and positive are stocks belonging to the same ETF. We report the score improvements of our setting over the counterpart for each evaluation metric.}
\caption{Ablation study on anchor selection for semantic alignment. We compare our setting (anchor: theme, positive: stock) with the counterpart (anchor: stock, positive: stock). We report the score improvements of our setting over the counterpart for each evaluation metric.}
\label{tab:ablation_anchor}
\setlength{\tabcolsep}{4.5pt} 
\renewcommand{\arraystretch}{0.9}
\vspace{-1em}
\resizebox{\columnwidth}{!}{
\begin{tabular}{l|ccc}
% \begin{tabular}{ll cccccc|cccccc}
\toprule
Model & \textbf{P@3} & \textbf{P@5} & \textbf{P@10} \\

\midrule
bge-small-en-v1.5~\cite{bge_embedding} & $\uparrow$ 0.1066 & $\uparrow$ 0.1433 & $\uparrow$ 0.1701 \\
bilingual-embedding-large~\cite{conneau2019unsupervised} & $\uparrow$ 0.1151 & $\uparrow$ 0.1227 & $\uparrow$ 0.1243 \\
multilingual-e5-large-instruct~\cite{wang2024multilingual} & $\uparrow$ 0.1375 & $\uparrow$ 0.1237 & $\uparrow$ 0.1361 \\
stella\_en\_1.5B\_v5~\cite{zhang2025jasperstelladistillationsota} & $\uparrow$ 0.1083 & $\uparrow$ 0.1340 & $\uparrow$ 0.1371 \\
gte-Qwen2-1.5B-instruct~\cite{li2023towards} & $\uparrow$ 0.1908 & $\uparrow$ 0.1907 & $\uparrow$ 0.1794 \\
SFR-Embedding-Mistral~\cite{SFRAIResearch2024} & $\uparrow$ 0.1202 & $\uparrow$ 0.1072 & $\uparrow$ 0.1181 \\
GritLM-7B~\cite{muennighoff2024generative} & $\uparrow$ \textbf{0.4261} & $\uparrow$ \textbf{0.4278} & $\uparrow$ \textbf{0.4155} \\
gte-Qwen2-7B-instruct~\cite{li2023towards} & $\uparrow$ 0.1168 & $\uparrow$ 0.1288 & $\uparrow$ 0.1258 \\
e5-mistral-7b-instruct~\cite{wang2023improving} & $\uparrow$ 0.0928 & $\uparrow$ 0.1093 & $\uparrow$ 0.1145 \\
Linq-Embed-Mistral~\cite{linq} & $\uparrow$ 0.0688 & $\uparrow$ 0.0649 & $\uparrow$ 0.0758 \\

\bottomrule
\end{tabular}
}
\end{table}

\subsubsection{Anchor Strategy: Stock–Stock vs. Theme-Based}

In this experiment, we compare two contrastive training configurations: Stock–Stock Alignment (SSA), where both the anchor and the positive are constituent stocks from the same ETF, and the theme-based anchor strategy, where the anchor is a textual description of the theme and the positive is a corresponding constituent stock. While SSA emphasizes local similarity within ETF holdings, the theme-based approach encourages alignment with a more abstract semantic concept.

As shown in Table~\ref{tab:ablation_anchor}, the theme-based anchor strategy consistently outperforms SSA across all models and evaluation thresholds. For instance, using \texttt{gte-Qwen2-1.5B}, P@3 increases from 0.2216 to 0.4124, and P@5 improves by 0.1907. The improvements are especially notable at lower $k$ values, indicating sharper ranking. These results suggest that using thematic descriptions as anchors offers more effective supervision than intra-stock alignment.

\subsubsection{Training Dataset: ETF-only vs. TRS}
Separately, we compare training on the Thematic ETF Set, which includes only real ETF data, to training on the more comprehensive TRS. While supervision from ETFs is grounded in actual investment products, it often shows a bias toward popular sectors like IT.
\begin{table}[htbp]
\small
\centering
\caption{Ablation study comparing the Thematic ETF Set (real ETFs only) with our broader TRS dataset. We report the score improvements of our setting over the counterpart for each evaluation metric.}
\label{tab:ablation_dataset}
\setlength{\tabcolsep}{4.5pt} 
\renewcommand{\arraystretch}{0.9}
\vspace{-1em}
\resizebox{\columnwidth}{!}{
\begin{tabular}{l|ccc}
\toprule
Model & \textbf{P@3} & \textbf{P@5} & \textbf{P@10} \\

\midrule
bge-small-en-v1.5~\cite{bge_embedding} & $\uparrow$ \textbf{0.0705} & $\uparrow$ \textbf{0.1186} & $\uparrow$ \textbf{0.1495} \\
bilingual-embedding-large~\cite{conneau2019unsupervised} & $\uparrow$ 0.0430 & $\uparrow$ 0.0485 & $\uparrow$ 0.0444 \\
multilingual-e5-large-instruct~\cite{wang2024multilingual} & $\uparrow$ \textbf{0.0705} & $\uparrow$ 0.0578 & $\uparrow$ 0.0598 \\
stella\_en\_1.5B\_v5~\cite{zhang2025jasperstelladistillationsota} & $\uparrow$ 0.0258 & $\uparrow$ 0.0216 & $\uparrow$ 0.0207 \\
gte-Qwen2-1.5B-instruct~\cite{li2023towards} & $\uparrow$ 0.0086 & $\uparrow$ 0.0113 & $\uparrow$ 0.0082 \\
SFR-Embedding-Mistral~\cite{SFRAIResearch2024} & $\uparrow$ 0.0515 & $\uparrow$ 0.0350 & $\uparrow$ 0.0382 \\
GritLM-7B~\cite{muennighoff2024generative} & $\uparrow$ 0.0327 & $\uparrow$ 0.0443 & $\uparrow$ 0.0433 \\
gte-Qwen2-7B-instruct~\cite{li2023towards} & $\uparrow$ 0.0670 & $\uparrow$ 0.0690 & $\uparrow$ 0.0479 \\
e5-mistral-7b-instruct~\cite{wang2023improving} & $\uparrow$ 0.0498 & $\uparrow$ 0.0485 & $\uparrow$ 0.0397 \\
Linq-Embed-Mistral~\cite{linq} & $\uparrow$ 0.0430 & $\uparrow$ 0.0299 & $\uparrow$ 0.0294 \\
\bottomrule
\end{tabular}
}
\end{table}

TRS addresses this by expanding theme coverage through sectoral classification and news-driven augmentation. 

As shown in Table~\ref{tab:ablation_dataset}, models trained on the TRS dataset consistently outperform their ETF-only counterparts across most settings. As an example, \texttt{bge-small-en-v1.5} gains +0.1186 in P@5, and \texttt{multilingual-e5-large-instruct} improves by +0.0705 in P@3. These findings highlight the importance of broad and balanced thematic coverage in contrastive learning for stock retrieval.

\section{Discussion}

The results from our experiments demonstrate not only the modeling strengths of THEME, but also its practical value in real-world thematic investing contexts. Rather than focusing solely on academic benchmarks, we highlight how the system’s design and performance translate into tangible benefits for various actors in the investment ecosystem.

First, the retrieval gains shown in Experiment \ref{sec:experiment1} support the use of THEME as an idea generation engine. Asset managers can input a high-level thematic description and retrieve a curated list of stocks that extends beyond traditional sector classifications. This facilitates the discovery of emerging or under-recognized companies that may not yet appear in static ETF portfolios. Since our approach consistently outperforms both LLM-based and rule-based retrieval baselines, it offers a reliable starting point for thematic universe construction across discretionary or quantitative workflows.

Second, Experiment \ref{sec:experiment2} shows that retrieved stocks are not only thematically relevant but also yield stronger short-term returns when combined with recent performance signals. This confirms the viability of THEME as a component in active investment processes. For example, portfolio managers can use it to periodically update allocations based on a combination of theme relevance and market momentum. Unlike traditional ETF reconstitution, this allows for more timely adjustments while maintaining thematic intent. The integration of semantic and temporal signals makes the system adaptable to both structural and market-driven shifts.

Third, our ablation study in Experiment \ref{sec:experiment3} emphasizes two design decisions that directly impact real-world performance: the choice of semantic anchors and the composition of the training dataset. We find that training with abstract theme descriptions as anchors yields better generalization than stock-to-stock alignment. In parallel, training on the expanded TRS leads to consistent retrieval gains over using real ETF data alone. This is especially important for practitioners seeking to cover a wide array of investment themes, including those not yet captured in the current ETF landscape. TRS also enables multi-theme associations, ensuring that stocks with diverse business models are not artificially constrained to a single category.

THEME is implemented as a modular, cloud-native platform with API-based access, making it deployable across a range of enterprise environments. Outputs can be integrated into discretionary workflows (e.g., analyst platforms), systematic engines (e.g., ranking or weighting modules), or even client-facing tools for personalized investing. For instance, a retail platform could allow users to enter a theme like "climate-resilient agriculture" and receive a dynamically updated portfolio reflecting both long-term relevance and short-term return potential.

Looking forward, we are incorporating real-time data sources, such as ESG events, earnings calls, and patent filings, to support additional use cases like intraday thematic scoring or event-driven rebalancing. Rather than replacing existing approaches, THEME is designed to serve as a thematic intelligence layer that augments decision-making for both passive and active strategies.

\section{Conclusion}

This work makes three key contributions to thematic investing.  
First, we propose THEME, a hierarchical contrastive learning framework that integrates semantic and temporal supervision to produce theme-aware stock embeddings.  
Second, we construct the TRS, a scalable dataset that addresses the coverage limitations of real-world thematic ETFs by incorporating industry classifications and financial news.  
Third, we demonstrate through extensive experiments that these design choices lead to consistent improvements in retrieval accuracy and portfolio performance, yielding a practical framework where embeddings can be pre-computed and stored for on-demand use.

THEME addresses a central challenge in thematic investing: finding stocks that are both conceptually aligned and responsive to market conditions. 
By combining thematic descriptions with recent return patterns, the system generates representations that support both interpretability and quantitative execution.

In our experiments, THEME outperforms extensive baselines including strong LLM and embedding-based baselines in retrieval precision and hit rate (Experiment \ref{sec:experiment1}), constructs portfolios with improved Sharpe ratios and cumulative returns (Experiment \ref{sec:experiment2}), and benefits from semantic anchoring and TRS-based training over conventional alternatives (Experiment \ref{sec:experiment3}). These results confirm that integrating semantic alignment with temporal sensitivity improves the relevance and profitability of thematic stock selection.

Beyond performance, THEME offers practical utility across investment workflows.  
Its outputs support thematic universe construction, dynamic portfolio rebalancing, and personalized investing.  
The resulting theme-aware embeddings are structured for efficient retrieval, enabling their integration into discretionary platforms, systematic engines, or client-facing applications.

While our framework performs well across metrics, several directions remain for future enhancement.  
Our supervision relies on curated ETF constituents, which may introduce theme bias; incorporating crowd-sourced labels or alternative benchmarks could improve robustness.  
We also plan to extend the system with richer inputs such as fundamentals, ESG signals, or supply chain structures.  
Enabling real-time inference based on live news or earnings events would further support intraday applications such as theme-aware risk monitoring or event-driven rebalancing.

In essence, THEME shows how semantic understanding and temporal modeling can be jointly applied to build more adaptive, scalable, and insightful systems for navigating complex investment themes in modern financial markets.

\section*{Acknowledgments}
This work was supported by National Research Foundation of Korea (NRF) grant (No. NRF-2022R1I1A4069163) and Institute of Information \& Communications Technology Planning \& Evaluation(IITP) grant (No. RS-2020-II201336, Artificial Intelligence Graduate School Program(UNIST)) funded by the Korea government(MSIT).

\newpage
\section*{GenAI Usage Disclosure}
The authors used EXAONE (LG AI Research) solely for language editing purposes, 
including grammar correction and improvements in readability. 
No substantive content such as text, data, figures, code, or novel interpretations 
was generated by the tool. 
All research ideas, methodology, experiments, analyses, and conclusions are entirely 
the work of the authors, who bear full responsibility for the accuracy and integrity 
of the manuscript.

% \clearpage

\bibliographystyle{ACM-Reference-Format}
\bibliography{my_bibliography}

\end{document}